\documentclass[11pt,a4paper]{article}
\pdfoutput=1
\usepackage{jheppub,braket,bm}
\allowdisplaybreaks[4]

\title{Light Neutralino Dark Matter Scenario in Supersymmetric Four-Higgs Doublet Model}
\author{Hidetoshi Kawase}
\affiliation{Department of Physics, Nagoya University, Nagoya 464-8602, Japan}
\emailAdd{hkawase@eken.phys.nagoya-u.ac.jp}

\abstract{
 We examine a possibility to explain the signals of dark matter direct detection
 from DAMA and CoGeNT experiments, under the framework of supersymmetric extension
 of the standard model.
 For this purpose, we introduce four Higgs doublet fields and assume that one pair of
 Higgs fields which have very small vacuum expectation values gives sizable effects
 for annihilation and scattering processes of dark matter.
 We show that the preferred parameter regions for DAMA and CoGeNT results can be
 simultaneously explained by supersymmetric four-Higgs doublet model with the
 parameters consistent with the observed value of dark matter relic abundance.
 The extra Higgs fields introduced as an explanation of the light dark matter
 scenario also explain the $Wjj$ anomaly reported by CDF.
}
\begin{document}
\maketitle
\section{Introduction}
The standard model is very successful in describing the interactions of elementary
particles.
However, from a cosmological point of view, we have not reached the theory which
can explain the whole nature of the universe.
In particular, a number of observations suggest that most of the mass in the
universe is composed of non-baryonic dark matter.
Interestingly, some experiments claim that they observed a signal compatible with
the light dark matter candidate by the direct detection search.
Among them, DAMA \cite{Bernabei:2008yi} and CoGeNT \cite{Aalseth:2010vx} experiments
have reported an annual modulation signals which support the existence of dark matter.
Moreover, CRESST experiment recently reported the observation of the signal events
with more than 4$\sigma$ significance \cite{Angloher:2011uu}.
Although other experiments such as XENON100 \cite{Aprile:2011hi} and CDMS-II
\cite{Ahmed:2010wy} have already reported the null result which exclude the preferred
region of the DAMA and CoGeNT results,
we put emphasis on the fact that there is 8.9$\sigma$ signature of an
annual modulation by DAMA experiment \cite{Bernabei:2010mq}.
For the validity of the analysis of DAMA experiment, some discussions are found
\cite{Ralston:2010bd}\cite{Nygren:2011xu}\cite{Blum:2011jf}
and it should be investigated more carefully in the future analysis.

Our goal in this paper is to construct a model which is capable of explaining
the light dark matter signals claimed by DAMA and CoGeNT experiments.
There are a number of works which investigate the model to explain these results
\cite{Foot:2008nw}\cite{Khlopov:2008ki}\cite{Masso:2009mu}\cite{Chang:2008gd}\cite{Cui:2009xq}\cite{Kim:2009ke}\cite{Bae:2010hr}\cite{Feng:2011vu}\cite{Gao:2011ka},
and we try to construct a concrete model as an extension of the standard model.
For this purpose, we need to introduce a stable neutral field which has a coupling
large enough to generate the observed rate of dark matter direct detection.
In other words, the key ingredient for the model building is a coupling between the
dark matter candidate particle and the field which mediate the dark matter-nucleus
scattering process.

As a model to realize a dark matter candidate, the minimal supersymmetric standard
model (MSSM) is well motivated because the lightest supersymmetric particle (LSP) is
stable by virtue of $R$-parity.
In fact, there are some attempts to interpret the DAMA result under the framework of
the MSSM \cite{Bottino:2002ry}\cite{Bottino:2003iu}\cite{Fornengo:2010mk}.
According to these study, large Yukawa coupling and light pseudoscalar
which mediate for the process of neutralino pair annihilation are required
to obtain the realistic value of neutralino relic abundance.
However, such scenario with large $\tan\beta$ and small pseudoscalar mass $M_A$ is
severely constrained by the direct searches at the Tevatron and the LHC
\cite{Feldman:2010ke}\cite{Kuflik:2010ah}\cite{Vasquez:2010ru}\cite{Vasquez:2011yq}.

With this knowledge, we consider a model of the extension of the MSSM which has
an extra pair of Higgs doublet superfields.
There are already some works which introduce multi-Higgs fields as an extension
of the MSSM
\cite{Escudero:2005hk}\cite{Gupta:2009wn}\cite{Kanemura:2010pa}\cite{Marshall:2010qi}\cite{Haba:2011ra}\cite{Aoki:2011yy}\cite{Kanemura:2011fy}
and we try to construct a model of light dark matter under the framework of
supersymmetric four-Higgs doublet model.
That is, if the LSP neutralino include some contribution from the extra higgsino
components, this lightest neutralino couples to the scalar components of the extra
Higgs fields.
Therefore, we can obtain a sizable contribution for the annihilation and scattering
process of dark matter and quarks if there are large Yukawa couplings for the
quarks and extra Higgs fields.
Note that we can achieve large Yukawa couplings for extra Higgs fields
assuming that these fields obtain quite small vacuum expectation values (VEVs)
and evade the usual Higgs search constraints.

Although it is not difficult to construct a model which is suitable for the
favored parameter region of DAMA experiment, reconciling the contradiction
with the null experiments such as XENON100 and CDMS-II is problematic.
Moreover, there is some discrepancy between the favored parameter regions of
DAMA and CoGeNT experiments.
Since the favored region for DAMA result is sensitive to the quenching factor of
sodium $Q_{\text{Na}}$, there are some arguments for the choice of this quenching
factor.
If we allow larger value of quenching factor than the default value
$Q_{Na} = 0.3 \pm 0.03$, DAMA favored region extend to include the smaller mass
range which relax the discrepancy with the CoGeNT result
\cite{Hooper:2010uy}\cite{Schwetz:2011xm}.

As the possible solution for the contradiction among the experiments,
there exist some works which propose an isospin violation in the dark matter-nucleus
cross section \cite{Feng:2011vu}\cite{Gao:2011ka}.
In fact, taking the ratio of neutron to proton couplings
$\lambda_n/\lambda_p \sim - 0.7$ and requiring large sodium quenching factor
$Q_{\text{Na}}$, DAMA and CoGeNT preferred region and the null result from XENON100 become
consistent for a dark matter with mass
$m_{\chi} \sim 8\,\mathrm{GeV}$ \cite{Hooper:2010uy}\cite{Schwetz:2011xm}.
In our model, we can easily achieve this isospin violation by adjusting the
extra Yukawa couplings for up-quark and down-quark as suggested in \cite{Gao:2011ka}.
Since it is still impossible to reconcile these results with the limit from
the CDMS experiment, we cannot claim the success for the explanation of all
experiments.
Even so, it is interesting that we can explain other experiments by the simple
extension of the MSSM, and the future experiments should test the model by
settling the contradiction.

In our model, we should have extra Higgs fields with mass of
$\mathcal{O}(100\,\mathrm{GeV})$ to acquire the dark matter-nucleus cross section
required to explain the DAMA and CoGeNT results.
Therefore, it may be possible to detect this extra Higgs field produced by the
collider experiments.
Interestingly, CDF Collaboration have reported an data on the dijet mass distribution
of the $W + j\,j$ channel which indicate a 3.2$\sigma$ excess around 150 GeV
\cite{Aaltonen:2011mk}.
One way to interpret this excess is to introduce new particles which couple to
quarks at tree-level \cite{Segre:2011tn}\cite{Cao:2011yt} and the extra Higgs fields
we introduce might be a solution for this anomaly.
Although this anomaly is somewhat questionable because it is not confirmed by D0
experiment, the connection with the dark matter search predicted by our model is
interesting and should be tested by LHC.

The outline of this paper is as follows.
In Section 2 we introduce the basic setup of a model of dark matter which is based on
supersymmetric four-Higgs doublet model.
In Section 3 we consider the thermal relic abundance of dark matter in our model
with the neutralino LSP scenario and compute the direct detection rate of dark
matter.
In Section 4 we investigate the constraints and discovery potential from other
experiments.
Our conclusion is given in Section 5.
\section{Supersymmetric four-Higgs doublet model}
Supersymmetric four-Higgs doublet model is a simple extension of the MSSM
by extra Higgs doublet superfields which have the same quantum numbers as
the original up-type Higgs $H_u$ and down-type Higgs $H_d$.
In this section, we illustrate the detail of the model which satisfy the
requirements for explaining the properties of dark matter.

To describe Higgs fields in this model, we use a following notation for
two pairs of up-type and down-type Higgs superfields:
\begin{equation}
 H_{ui} = \binom{H_{ui}^+}{H_{ui}^0},\qquad
  H_{di} = \binom{H_{di}^0}{H_{di}^-},\qquad (i = 1,2),
\end{equation}
where we assume that the masses of quarks and leptons come mainly from
the VEVs of $H_{u1}$ and $H_{d1}$.
In other words, we introduce $H_{u2}$ and $H_{d2}$ as fields with very small VEVs
compared with $H_{u1}$ and $H_{d1}$.
From now on, we will refer to the Higgs doublets $H_{u1}$, $H_{d1}$ as ``original Higgs''
which correspond to the fields originally contained in the MSSM
and the additional Higgs doublets $H_{u2}$, $H_{d2}$ as ``extra Higgs''.

Now we can write the superpotential terms involving these fields as
\begin{equation}
 W = \sum_{i,j=1,2}\mu_{ij}H_{ui}H_{dj} + \sum_{i=1,2}
  \left[(Y_{ui})_{ab}Q^a\bar{u}^bH_{ui}
  - (Y_{di})_{ab}Q^a\bar{d}^bH_{di} - (Y_{ei})_{ab}L^a\bar{e}^bH_{di}\right]
  \label{eq:sp}
\end{equation}
where $a$, $b$ ($= 1,2,3$) are flavor indices for the matter superfields.
Other terms such as $L^aH_{ui}$ are assumed to be forbidden by $R$-parity.
Once Higgs fields obtain the VEVs which break the electroweak symmetry,
each field contributes to the masses of quarks and leptons.
Since we assume that $H_{u2}$ and $H_{d2}$ obtain very small VEVs
(of order less than a few MeV), they do not give the large
contribution to the fermion masses even if we assign $\mathcal{O}(1)$ couplings
for the Yukawa interactions among the matter fields and extra Higgs fields.
This assumption is crucial to explain the results of dark matter search
as we will show in the next section.

The soft supersymmetry breaking terms which are relevant to the Higgs scalar
potential are
\begin{equation}
 V_{\text{soft}} = \sum_{i=1,2}m_{ui}^2|H_{ui}|^2 + \sum_{i=1,2}m_{di}^2|H_{di}|^2
  + \sum_{i,j=1,2}(b_{ij}H_{ui}H_{dj} + \text{c.c.}).
\end{equation}
Then the scalar potential of Higgs fields can be written as
\begin{align}
 V &= \sum_{i=1,2}
 \left[
 (|\mu_{i1}|^2 + |\mu_{i2}|^2 + m_{ui}^2)(|H_{ui}^0|^2 + |H_{ui}^+|^2)
 + (|\mu_{1i}|^2 + |\mu_{2i}|^2 + m_{di}^2)(|H_{di}^0|^2 + |H_{di}^-|^2)
 \right] \notag \\
 &\qquad + [(\mu_{11}^*\mu_{21} + \mu_{12}^*\mu_{22})
 (H_{u1}^{0*}H_{u2}^0 + H_{u1}^{+*}H_{u2}^+) \notag \\
 &\qquad\qquad + (\mu_{11}^*\mu_{12} + \mu_{21}^*\mu_{22})
 (H_{d1}^{0*}H_{d2}^0 + H_{d1}^{-*}H_{d2}^-) + \text{c.c.}] \notag \\
 &\qquad +
 \left[\sum_{i,j=1,2}b_{ij}(H_{ui}^+H_{dj}^- - H_{ui}^0H_{dj}^0) + \text{c.c.}\right]
 \notag \\
 &\qquad + \frac{g^2 + g^{\prime 2}}{8}
 \left[\sum_{i=1,2}
 \left(|H_{ui}^0|^2 + |H_{ui}^+|^2 - |H_{di}^0|^2 + |H_{di}^-|^2\right)\right]^2 \notag \\
 &\qquad + \frac{g^2}{2}\left[|(H_{ui}^{+*}H_{ui}^0 + H_{di}^{0*}H_{di}^-)|^2
 - (|H_{ui}^0|^2 - |H_{di}^0|^2)(|H_{uj}^+|^2 - |H_{dj}^-|^2)\right].
\end{align}
Although the couplings $\mu_{ij}$ and $b_{ij}$ are complex parameters in general,
from now on we take these parameters real for the sake of simplicity.

As was pointed out in \cite{Masip:1995sm}, the VEVs of Higgs fields in this model is
always real when all of the parameters in the Higgs potential are real at tree-level.
Now we assume that neutral Higgs fields obtain real VEVs
$\braket{H_{ui}^0} = v_{ui} \neq 0$, $\braket{H_{di}^0} = v_{di} \neq 0$
and the VEVs of charged Higgs fields vanish so as not to break the electromagnetic
symmetry.
So we can parameterize these VEVs as follows:
\begin{equation}
 v_{u1} = v_1\sin\beta_1,\quad v_{d1} = v_1\cos\beta_1,\quad
  v_{u2} = v_1\tan\omega\sin\beta_2,\quad v_{d2} = v_1\tan\omega\cos\beta_2
\end{equation}
where $v^2 \equiv v_1^2(1 + \tan^2\omega) = 2M_Z^2/(g^2 + g^{\prime 2})$.
Since we consider the situation where the extra Higgs fields obtain very small
VEVs, we put $\tan\omega \ll 1$.

Demanding that the first derivatives of the Higgs potential with respect to
the neutral Higgs vanish,
$\partial V/\partial H_{ui}^0 = \partial V/\partial H_{di}^0 = 0$, we obtain
\begin{align}
 0 &= \mu_{11}^2 + \mu_{12}^2 + m_{u1}^2 
 + (\mu_{11}\mu_{21} + \mu_{12}\mu_{22})\tan\omega\sin^{-1}\beta_1
 \sin\beta_2 \notag \\
 &\qquad - b_{11}\tan^{-1}\beta_1 - b_{12}\tan\omega\sin^{-1}\beta_1\cos\beta_2
 \notag \\
 &\qquad + \frac{M_Z^2}{2(1 + \tan^2\omega)}\left[\sin^2\beta_1 - \cos^2\beta_1
 + \tan^2\omega(\sin^2\beta_2 - \cos^2\beta_2)\right], \label{eq:pot1} \\
 0 &= \mu_{11}^2 + \mu_{21}^2 + m_{d1}^2
 + (\mu_{11}\mu_{12} + \mu_{21}\mu_{22})\tan\omega\cos^{-1}\beta_1\cos\beta_2 \notag \\
 &\qquad - b_{11}\tan\beta_1 - b_{21}\tan\omega\cos^{-1}\beta_1\sin\beta_2 \notag \\
 &\qquad - \frac{M_Z^2}{2(1 + \tan^2\omega)}\left[\sin^2\beta_1 - \cos^2\beta_1
 + \tan^2\omega(\sin^2\beta_2 - \cos^2\beta_2)\right], \label{eq:pot2} \\
 0 &= \mu_{21}^2 + \mu_{22}^2 + m_{u2}^2
 + (\mu_{11}\mu_{21} + \mu_{12}\mu_{22})\tan^{-1}\omega\sin\beta_1\sin^{-1}\beta_2
 \notag \\
 &\qquad - b_{22}\tan^{-1}\beta_2 - b_{21}\tan^{-1}\omega\cos\beta_1\sin^{-1}\beta_2
 \notag \\
 &\qquad + \frac{M_Z^2}{2(1 + \tan^2\omega)}\left[\sin^2\beta_1 - \cos^2\beta_1
 + \tan^2\omega(\sin^2\beta_2 - \cos^2\beta_2)\right], \label{eq:pot3} \\
 0 &= \mu_{12}^2 + \mu_{22}^2 + m_{d2}^2
 + (\mu_{11}\mu_{12} + \mu_{21}\mu_{22})\tan^{-1}\omega\cos\beta_1\cos^{-1}\beta_2
 \notag \\
 &\qquad - b_{22}\tan\beta_2 - b_{12}\tan^{-1}\omega\sin\beta_1\cos^{-1}\beta_2
 \notag \\
 &\qquad - \frac{M_Z^2}{2(1 + \tan^2\omega)}\left[\sin^2\beta_1 - \cos^2\beta_1
 + \tan^2\omega(\sin^2\beta_2 - \cos^2\beta_2)\right]. \label{eq:pot4}
\end{align}
For $\tan\omega \ll 1$, equations \eqref{eq:pot1} and \eqref{eq:pot2} reduce to the
approximate constraints
\begin{align}
 0&\simeq \mu_{11}^2 + \mu_{12}^2 + m_{u1}^2 - b_{11}\tan^{-1}\beta_1
 - \frac{M_Z^2}{2}\cos 2\beta_1, \label{eq:const1} \\
 0&\simeq \mu_{11}^2 + \mu_{21}^2 + m_{d1}^2 - b_{11}\tan\beta_1
 + \frac{M_Z^2}{2}\cos 2\beta_1, \label{eq:const2}
\end{align}
and to satisfy the constraints \eqref{eq:pot3} and \eqref{eq:pot4},
there are four possible choices of the parameters, i.e.,
\begin{enumerate}
 \item $b_{12}, b_{21} \simeq 0$ and $\mu_{11}, \mu_{22} \simeq 0$.
 \item $b_{12}, b_{21} \simeq 0$ and $\mu_{12}, \mu_{21} \simeq 0$.
 \item $b_{12}, b_{21} \simeq 0$
       and $\mu_{11} \simeq \pm \mu_{22}$, $\mu_{12} \simeq \mp \mu_{21}$.
 \item $b_{12} \simeq (\mu_{11}\mu_{12} + \mu_{21}\mu_{22})\tan^{-1}\beta_1$ and
       $b_{21} \simeq (\mu_{11}\mu_{21} + \mu_{12}\mu_{22})\tan\beta_1$,
\end{enumerate}
because of large $\tan^{-1}\omega$.
Since we need a tuning of the parameters for the case 3 and 4,
we will adopt the case 1 and 2 for the subsequent analysis.

Now let us consider the masses of the particles related with the Higgs sector.
Since neutralinos contain the higgsino components corresponding to the extra
Higgs, we write the masses of neutralinos in the basis
$\psi^0 = (\tilde{B},\tilde{W}^3,\tilde{H}_{d1}^0,\tilde{H}_{u1}^0,\tilde{H}_{d2}^0,
\tilde{H}_{u2}^0)$ as
\begin{equation}
 M_{\tilde{N}} \simeq
 \begin{pmatrix}
  M_1 & 0 & -c_{\beta}s_W M_Z & s_{\beta}s_W M_Z & 0 & 0 \\
  0 & M_2 & c_{\beta}c_W M_Z & -s_{\beta}c_W M_Z & 0 & 0 \\
  -c_{\beta}s_W M_Z & c_{\beta}c_W M_Z & 0 & -\mu_{11} & 0 & -\mu_{21} \\
  s_{\beta}s_W M_Z & -s_{\beta}c_W M_Z & -\mu_{11} & 0 & -\mu_{12} & 0 \\
  0 & 0 & 0 & -\mu_{12} & 0 & -\mu_{22} \\
  0 & 0 & -\mu_{21} & 0 & -\mu_{22} & 0
 \end{pmatrix}
\end{equation}
where we adopt the notation that
\begin{equation}
 s_{\beta} \equiv \sin\beta_1,\qquad c_{\beta} \equiv \cos\beta_1,\qquad
  t_{\beta} \equiv \tan\beta_1
\end{equation}
and neglect the elements which are proportional to $\tan\omega$.
Similarly, the mass matrix of charginos in the basis
$\psi = (\tilde{W}^+,\tilde{H}_{u1}^+,\tilde{H}_{u2}^+,\tilde{W}^-,\tilde{H}_{d1}^-,
\tilde{H}_{d2}^-)$ is
\begin{equation}
 M_{\tilde{C}} =
  \begin{pmatrix}
   0 & \mathcal{M}^T \\
   \mathcal{M} & 0
  \end{pmatrix}
  ,\qquad \mathcal{M} \simeq
  \begin{pmatrix}
   M_2 & \sqrt{2}s_{\beta}M_W & 0 \\
   \sqrt{2}c_{\beta}M_W & \mu_{11} & \mu_{21} \\
   0 & \mu_{12} & \mu_{22}
  \end{pmatrix}
\end{equation}
and the spectrum of these particles can be modified from that of the MSSM if
$\mu_{12}$ and $\mu_{21}$ are not so small.

Next let us consider the Higgs scalar fields.
For the real parts of the neutral Higgs fields
$(\mathrm{Re}H_{u1}^0,\mathrm{Re}H_{d1}^0,\mathrm{Re}H_{u2}^0,\mathrm{Re}H_{d2}^0)$,
we have CP-even Higgs mass matrix
\begin{equation}
 M_E^2 \simeq
  \begin{pmatrix}
   b_{11}t_{\beta}^{-1} + M_Z^2s_{\beta}^2 & -b_{11} - M_Z^2s_{\beta}c_{\beta} & 0 & 0 \\
   -b_{11} - M_Z^2s_{\beta}c_{\beta} & b_{11}t_{\beta} + M_Z^2c_{\beta}^2 & 0 & 0 \\
   0 & 0 & \Delta_u & -b_{22} \\
   0 & 0 & -b_{22} & \Delta_d
  \end{pmatrix}
  ,
\end{equation}
and for the imaginary parts of the neutral fields
$(\mathrm{Im}H_{u1}^0,\mathrm{Im}H_{d1}^0,\mathrm{Im}H_{u2}^0,\mathrm{Im}H_{d2}^0)$,
CP-odd Higgs mass matrix is
\begin{equation}
 M_O^2 \simeq
  \begin{pmatrix}
   b_{11}t_{\beta} & b_{11} & 0 & 0 \\
   b_{11} & b_{11}t_{\beta}^{-1} & 0 & 0 \\
   0 & 0 & \Delta_u & b_{22} \\
   0 & 0 & b_{22} & \Delta_d
  \end{pmatrix}
\end{equation}
where $\Delta_u$ and $\Delta_d$ are defined as
\begin{equation}
 \Delta_u \equiv \mu_{21}^2 + \mu_{22}^2 + m_{u2}^2 - \frac{M_Z^2}{2}c_{2\beta},
  \qquad
  \Delta_d \equiv \mu_{12}^2 + \mu_{22}^2 + m_{d2}^2 + \frac{M_Z^2}{2}c_{2\beta}.
\end{equation}
Here we used \eqref{eq:const1} and \eqref{eq:const2} to simplify the matrix elements,
and omitted the elements which are proportional to $b_{12}$, $b_{21}$ and $\tan\omega$.
It should be pointed out that these mass matrices are block diagonal and the structure
of the blocks corresponding to the original Higgs $H_{u1}^0$ and $H_{d1}^0$ are similar
to that of the MSSM.
Moreover, for the blocks corresponding to the extra Higgs fields, the mass eigenvalues
are degenerate between CP-even Higgs and CP-odd Higgs.
So we find the mass eigenstates of the neutral Higgs fields;
\begin{align}
 \binom{H_{u1}^0}{H_{d1}^0} &\simeq \binom{v_u}{v_d}
 + \frac{1}{\sqrt{2}}
 \begin{pmatrix}
  c_{\alpha} & s_{\alpha} \\
  -s_{\alpha} & c_{\alpha}
 \end{pmatrix}
 \binom{h^0}{H^0} + \frac{i}{\sqrt{2}}
 \begin{pmatrix}
  c_{\beta} & s_{\beta} \\
  s_{\beta} & -c_{\beta}
 \end{pmatrix}
 \binom{A^0}{G^0}, \\
 \binom{H_{u2}^0}{H_{d2}^0} &\simeq \frac{1}{\sqrt{2}}
 \begin{pmatrix}
  c_{\gamma} & s_{\gamma} \\
  -s_{\gamma} & c_{\gamma}
 \end{pmatrix}
 \binom{\phi_{E1}^0}{\phi_{E2}^0} + \frac{i}{\sqrt{2}}
 \begin{pmatrix}
  c_{\gamma} & -s_{\gamma} \\
  s_{\gamma} & c_{\gamma}
 \end{pmatrix}
 \binom{\phi_{O1}^0}{\phi_{O2}^0}
\end{align}
where $s_{\alpha} \equiv \sin\alpha$ and $c_{\alpha} \equiv \cos\alpha$ represent
the mixing of the real components of the original Higgs
and $s_{\gamma} \equiv \sin\gamma$ and $c_{\gamma} \equiv \cos\gamma$ are the
mixing angle of the extra Higgs.
Here we define the mixing angle $\gamma$ so that the eigenstates corresponding to
the lighter mass eigenvalue are $\phi_{E1}^0$ and $\phi_{O1}^0$.
Note that the sign of the mixing angles of the extra Higgs are opposite between
CP-even components and CP-odd components.

The mass matrix of charged Higgs fields
$(H_{u1}^{+*},H_{d1}^-,H_{u2}^{+*},H_{d2}^-)$
can be written in terms of that of CP-odd Higgs fields as
\begin{equation}
 M_+^2 \simeq M_O^2 + M_W^2
  \begin{pmatrix}
   c_{\beta}^2 & s_{\beta}c_{\beta} & 0 & 0 \\
   s_{\beta}c_{\beta} & s_{\beta}^2 & 0 & 0 \\
   0 & 0 & c_{2\beta} & 0 \\
   0 & 0 & 0 & -c_{2\beta}
  \end{pmatrix}
  .
\end{equation}
\section{Neutralino dark matter scenario}
In this section, we investigate the very light ($\sim 10\,\mathrm{GeV}$) neutralino
dark matter scenario under the framework of supersymmetric four-Higgs doublet model.
We expect that this light neutralino explain the signals of dark matter by means of
the exchange of extra Higgs fields for annihilation and scattering processes of
neutralino.
Especially, our model have an advantage that we can realize an isospin violation in
the dark matter-nucleus cross section which is required to satisfy both of the signals
of DAMA and CoGeNT \cite{Feng:2011vu}\cite{Gao:2011ka}.

To discuss the neutralino dark matter scenario in this model, we first investigate
the mixing structure of the lightest neutralino $\chi_1^0$.
This lightest neutralino is guaranteed to be stable by virtue of $R$-parity
as long as this particle is the LSP.
Let us write the lightest neutralino in terms of its components fields, bino
$\tilde{B}$, wino $\tilde{W}^3$ and higgsinos $\tilde{H}_{di}$, $\tilde{H}_{ui}$:
\begin{equation}
 \chi_1^0 = N_{11}\tilde{B} + N_{12}\tilde{W}^3 + N_{13}\tilde{H}_{d1}
  + N_{14}\tilde{H}_{u1} + N_{15}\tilde{H}_{d2} + N_{16}\tilde{H}_{u2}
\end{equation}
where $N$ is a unitary matrix which diagonalize the neutralino mass matrix,
$N^*M_{\tilde{N}}N^{-1} = M_{\tilde{N}}^{\text{diag}}$.
If we denote the mass of the lightest neutralino as $m_{\chi}$,
the coefficients $N_{1i}$ ($i=1,\dots,6$) should satisfy the relation
\begin{align}
 0 &= (M_1 - m_{\chi})N_{11} - c_{\beta}s_WM_ZN_{13} + s_{\beta}s_WM_ZN_{14},
 \label{eq:mix1} \\
 0 &= (M_2 - m_{\chi})N_{12} + c_{\beta}c_WM_ZN_{13} - s_{\beta}c_WM_ZN_{14}, \\
 0 &= -c_{\beta}s_WM_ZN_{11} + c_{\beta}c_WM_ZN_{12} - m_{\chi}N_{13}
 - \mu_{11}N_{14} - \mu_{21}N_{16}, \\
 0 &= s_{\beta}s_WM_ZN_{11} - s_{\beta}c_WM_ZN_{12} - \mu_{11}N_{13}
 - m_{\chi}N_{14} - \mu_{12}N_{15}, \\
 0 &= -\mu_{12}N_{14} - m_{\chi}N_{15} - \mu_{22}N_{16}, \\
 0 &= -\mu_{21}N_{13} - \mu_{22}N_{15} - m_{\chi}N_{16}. \label{eq:mix2}
\end{align}
as in the case of the MSSM \cite{Bottino:2008xc}.

Meanwhile, we are interested in the situation where
(1) $\mu_{11}, \mu_{22} \simeq 0$ or (2) $\mu_{12}, \mu_{21} \simeq 0$.
In the case 2, it is obvious that the lightest neutralino can contain
very small components of the extra higgsino $\tilde{H}_{d2}^0$ and $\tilde{H}_{u2}^0$.
Therefore, we adopt the case 1 and assume
$\mu_{11}, \mu_{22} \ll M_1 \ll M_Z, \mu_{12}, \mu_{21}, M_2$.
The assumption $M_1 \ll M_Z$ is to realize the situation of the very light
neutralino dark matter of $\mathcal{O}(10\,\text{GeV})$.
Since the masses of charginos are constrained by LEP II bound to be heavier than
about 100 GeV, the lightest neutralino cannot be wino-like or higgsino-like.
With this assumption, we can solve \eqref{eq:mix1}--\eqref{eq:mix2} to obtain
\begin{align}
 N_{12} &\simeq \frac{M_1 - m_{\chi}}{M_2}t_W^{-1}N_{11} \ll 1, \\
 N_{13} &\simeq -\frac{m_{\chi}}{\mu_{21}}N_{16} \ll 1, \\
 N_{14} &\simeq -\frac{m_{\chi}}{\mu_{12}}N_{15} \ll 1, \\
 N_{15} &\simeq s_{\beta}s_W\frac{M_Z}{\mu_{12}}N_{11}, \label{eq:n15} \\
 N_{16} &\simeq -c_{\beta}s_W\frac{M_Z}{\mu_{21}}N_{11}, \label{eq:n16}
\end{align}
and
\begin{equation}
 1 \simeq N_{11}^2 + N_{15}^2 + N_{16}^2 \simeq
  \left[1 + s_W^2M_Z^2
   \left(\frac{s_{\beta}^2}{\mu_{12}^2} + \frac{c_{\beta}^2}{\mu_{21}^2}\right)\right]
  N_{11}^2.
\end{equation}
The mass of the lightest neutralino can be written as
\begin{equation}
 m_{\chi} \simeq M_1\left(1 + \frac{s_W^2M_Z^2}{\mu_{12}\mu_{21}}\right)^{-1}.
\end{equation}
\subsection{Dark matter relic abundance}
For a stable neutralino, we can estimate the neutralino thermal relic abundance
by calculating the neutralino pair annihilation cross section.
In this model, it is reasonable to assume that the extra Higgs exchange contribution
is dominant for the process $\chi_1^0\,\chi_1^0 \to f\,\bar{f}$
if the Yukawa couplings for the extra Higgs given in \eqref{eq:sp} are large enough.
Especially, we will investigate the situation where the extra Higgs which
correspond to the lighter mass eigenvalue ($\phi_{E1}$ and $\phi_{O1}$) dominate
the contribution of the pair annihilation, and assume that the other eigenstates
($\phi_{E2}$ and $\phi_{O2}$) are heavy enough to be neglected.
Since CP-even extra Higgs $\phi_{E1}$ and CP-odd extra Higgs $\phi_{O1}$ are degenerate
as pointed out in the previous section, we denote their masses as $m_{\phi}$ from
now on.

Although the extra Higgs exchange contribution for the neutralino pair annihilation
process is generated from the Yukawa interactions in \eqref{eq:sp}, such contributions
for the flavor interactions are severely constrained by plenty of experiments.
First, if there is a large Yukawa coupling for the electron and extra Higgs,
it leads to the process $e^+\,e^- \to \phi_{E1}^0$ and the constraint
$m_{\phi} \gtrsim 200\,\text{GeV}$ comes from the LEP II experiment.
Moreover, off-diagonal components of the Yukawa matrix for the quarks and extra
Higgs are prohibited because such terms induce the flavor changing neutral current
at tree-level.
For the subsequent analysis, we simply assume that all Yukawa matrices for the
quark sector are proportional to a unit matrix in a basis of quark mass eigenstates;
\begin{equation}
 V_{uL}^{\sf T}Y_{u2}V_{uR} \simeq y_u^{\prime}\cdot\bm{1}
  ,\qquad V_{dL}^{\sf T}Y_{d2}V_{dR} \simeq y_d^{\prime}\cdot\bm{1},
  \label{eq:yukawa}
\end{equation}
where
\begin{equation}
 V_{uL}^{\sf T}Y_{u1}\braket{H_{u1}}V_{uR} \simeq \mathrm{diag}\,(m_u,m_c,m_t),\qquad
  V_{dL}^{\sf T}Y_{d1}\braket{H_{d1}}V_{dR} \simeq \mathrm{diag}\,(m_d,m_s,m_b)
\end{equation}
and we parameterize the size of extra Yukawa couplings by $y_u^{\prime}$ and
$y_d^{\prime}$.
To avoid the constraint from the extra Higgs production process by LEP II,
we simply assume that there is no Yukawa interaction term between leptons and extra
Higgs.

With this assumption, we calculate the process $\chi_1^0\,\chi_1^0 \to f\,\bar{f}$
where $f$ represents quark fields which contribute to the annihilation process.
Note that we neglect the quark masses and consider the contribution which comes
from the exchange of extra Higgs fields $\phi_{E1}$ and $\phi_{O1}$.
For the process with the final state up-type quarks,
$\chi_1^0\,\chi_1^0 \to u\,\bar{u}$,
we can compute the thermally averaged annihilation cross section
$\braket{\sigma v}_u$ which is given by
\begin{equation}
 \braket{\sigma v}_u \simeq \frac{3}{4\pi}\frac{m_{\chi}^2}{m_{\phi}^4}
  g^{\prime 2}y_u^{\prime 2}N_{11}^2c_{\gamma}^2
  (N_{15}s_{\gamma} - N_{16}c_{\gamma})^2.
\end{equation}
Here, we leave only an s-wave annihilation contribution because p-wave contribution
is suppressed by a factor $x^{-1} = T / m_{\chi}$.
Similarly, for the process involving down-type quarks,
$\chi_1^0\,\chi_1^0 \to d\,\bar{d}$,
we obtain
\begin{equation}
 \braket{\sigma v}_d \simeq \frac{3}{4\pi}\frac{m_{\chi}^2}{m_{\phi}^4}
  g^{\prime 2}y_d^{\prime 2}N_{11}^2s_{\gamma}^2
  (N_{15}s_{\gamma} - N_{16}c_{\gamma})^2.
\end{equation}
Therefore they sum up the result
\begin{equation}
 \braket{\sigma v} = N_u\cdot\braket{\sigma v}_u + N_d\cdot\braket{\sigma v}_d
  = \frac{3}{4\pi}\frac{m_{\chi}^2}{m_{\phi}^4}g^{\prime 2}
  N_{11}^2(2y_u^{\prime 2}c_{\gamma}^2 + 3y_d^{\prime 2}s_{\gamma}^2)
  (N_{15}s_{\gamma} - N_{16}c_{\gamma})^2
  \label{eq:sigmav}
\end{equation}
where we take $N_u = 2$ and $N_d = 3$ because quark fields other than top quark
give the contribution for this annihilation process.
Since the annihilation cross section is proportional to $m_{\chi}^2/m_{\phi}^4$,
this cross section tends to be small for the very light neutralino.

Using the annihilation cross section, the relic abundance of the neutralino can be
obtained as \cite{Kolb:1990vq}
\begin{equation}
 \Omega_{\chi}h^2 \simeq
  \frac{x_f \cdot 1.07 \times 10^9 (\mathrm{GeV}^{-1})}
  {g_*^{1/2}m_{\text{Pl}}\braket{\sigma v}}
\end{equation}
where $x_f$ is a freeze-out temperature and $g_*$ is an effective massless degree
of freedom.
Now we can determine $y_f^{\prime}$ from the experimental value
$\Omega_{\chi}h^2 \simeq 0.1$.
For the case $|y_d^{\prime}| \gtrsim |y_u^{\prime}|$, required value of $y_d^{\prime}$
can be estimated to be
\begin{equation}
 |y_d^{\prime}| \gtrsim 2.2
  \cdot \left(\frac{0.1}{\Omega_{\chi}h^2}\right)^{1/2}
  \cdot \left(\frac{8\,\mathrm{GeV}}{m_{\chi}}\right)
  \cdot \left(\frac{m_{\phi}}{150\,\mathrm{GeV}}\right)^2
  \cdot \left(\frac{0.92}{N_{11}}\right)^2,
\end{equation}
where we use the values $g_* \simeq 8$ and $x_f \simeq 20$.
Note that this Yukawa coupling tends to blow up above the mass scale $m_{\phi}$
due to the contribution to the renormalization group equation from the loop
mediated by the extra Higgs.
So this imply that there might be some other new physics around a few TeV.
But we will not discuss further about this topic in this paper.
\subsection{Direct detection cross section}
As we mentioned in the introduction, direct detection experiment is interesting because
DAMA and CoGeNT experiments have reported that they observed the annual
modulation of the signals which is compatible with a light dark matter scenario.
So we investigate whether our model can explain these signals by computing the
rate of the scattering between dark matter and nucleus.
Requiring that the proper value of the relic abundance of dark matter
$\Omega_{\chi}h^2 \simeq 0.1$ is generated, we can expect to derive some prediction for
the direct detection rate.
For the explanation of the results of DAMA and CoGeNT, an isospin violation plays
the crucial role as we show later.

To calculate the detection rate of neutralino-nucleus elastic scattering,
let us consider the effective Lagrangian which describe the interaction between
neutralino and quarks.
If we consider only the extra Higgs exchange contribution, the effective Lagrangian
reads
\begin{align}
 \mathcal{L}_{\text{eff}} = \mathcal{A}_1^f(\bar{\chi}\chi)(\bar{f}f)
 + \mathcal{A}_2^f(\bar{\chi}\gamma^5\chi)(\bar{f}\gamma^5f)
\end{align}
where
\begin{align}
 \mathcal{A}_1^u &= \frac{\sqrt{2}g^{\prime}y_u^{\prime}}{4m_{\phi}^2}N_{11}c_{\gamma}
 (N_{15}s_{\gamma} + N_{16}c_{\gamma}), \\
 \mathcal{A}_2^u &= -\frac{\sqrt{2}g^{\prime}y_u^{\prime}}{4m_{\phi}^2}N_{11}c_{\gamma}
 (N_{15}s_{\gamma} - N_{16}c_{\gamma}), \\
 \mathcal{A}_1^d &= -\frac{\sqrt{2}g^{\prime}y_d^{\prime}}{4m_{\phi}^2}N_{11}s_{\gamma}
 (N_{15}s_{\gamma} + N_{16}c_{\gamma}), \\
 \mathcal{A}_2^d &= -\frac{\sqrt{2}g^{\prime}y_d^{\prime}}{4m_{\phi}^2}N_{11}s_{\gamma}
 (N_{15}s_{\gamma} - N_{16}c_{\gamma}).
\end{align}
Note that $\mathcal{A}_1^{u/d}$ comes from CP-even extra Higgs exchange and
$\mathcal{A}_2^{u/d}$ comes from CP-odd extra Higgs exchange.
Since the term with coefficient $\mathcal{A}_2^{u/d}$ give the velocity-dependent
contribution to the elastic scattering, only the term with $\mathcal{A}_1^{u/d}$ are
important for our analysis.

The spin-independent cross section with a nucleus with $Z$ protons and $A - Z$
neutrons is given by
\begin{equation}
 \sigma_0^{\text{SI}} = \frac{4\mu_{\chi}^2}{\pi}
  (\lambda_pZ + \lambda_n(A - Z))^2
\end{equation}
where $\lambda_N$ ($N = p,n$) is four-fermion coupling of the neutralino with
point like nucleus and $\mu_{\chi} = m_{\chi}m_A / (m_{\chi} + m_A)$ is the
reduced neutralino-nucleus mass.
Now we can obtain $\lambda_N$ from the parton level couplings as
\begin{align}
 \lambda_N &= \braket{N|\bar{u}u|N}\mathcal{A}_1^u
 + \braket{N|\bar{d}d|N}\mathcal{A}_1^d \notag \\
 &= \frac{\sqrt{2}g^{\prime}}{4m_{\phi}^2}N_{11}
 \left(\frac{m_N}{m_u}f_u^Ny_u^{\prime}c_{\gamma}
 - \frac{m_N}{m_d}f_d^Ny_d^{\prime}s_{\gamma}\right)
 (N_{15}s_{\gamma} + N_{16}c_{\gamma})
 \label{eq:lambdan}
\end{align}
where 
\begin{equation}
 \braket{N|\bar{f}f|N} = \frac{m_N}{m_f}f_f^N,
\end{equation}
and we consider the contribution only from up-quark and down-quark because the
contributions from other quarks are suppressed by their masses.\footnote{
The contribution from strange quark is possibly large if $f_s^N$ is much
larger than $f_u^N$ and $f_d^N$.
But we adopt the small value of $f_s^N$ as suggested by the lattice study
\cite{Ohki:2008ff}.
}
The parameters $f_u^N$ and $f_d^N$ are obtained as
\cite{Pavan:2001wz}\cite{Belanger:2008sj}
\begin{equation}
 f_u^p = 0.023,\qquad f_d^p = 0.033,\qquad
 f_u^n = 0.018,\qquad f_d^n = 0.042.
\end{equation}

Since we can take arbitrary values for the Yukawa couplings of extra Higgs
($y_u^{\prime}$ and $y_d^{\prime}$) independent of the masses of quarks, the effective
coupling $\lambda_N$ are generally different between proton and neutron.
In other words, we can make large isospin violation for the spin-independent
scattering as long as $f_f^N$ have different values each other
\cite{Feng:2011vu}\cite{Gao:2011ka}.
Now we parameterize the isospin violation by the ratio
$r \equiv \lambda_n / \lambda_p$ and obtain
\begin{equation}
 y_d^{\prime} = k\frac{m_d}{m_u}\tan^{-1}\gamma \cdot y_u^{\prime},\qquad
  k \equiv \frac{f_u^n - rf_u^p}{f_d^n - rf_d^p}.
  \label{eq:yprime}
\end{equation}
In Figure 1, we show the contour plot of $y_d^{\prime} / y_u^{\prime}$ for various
values of $r$ and $\gamma$ with the quark mass ratio $m_u/m_d \simeq 0.55$.
Using \eqref{eq:yprime}, we can rewrite \eqref{eq:lambdan} as
\begin{equation}
 \lambda_N = \frac{\sqrt{2}g^{\prime}}{4m_{\phi}^2}N_{11}\frac{m_N}{m_u}
  (f_u^N - kf_d^N)y_u^{\prime}c_{\gamma}
  (N_{15}s_{\gamma} + N_{16}c_{\gamma}).
\end{equation}

\begin{figure}[t]
\begin{center}
\includegraphics[width=10cm]{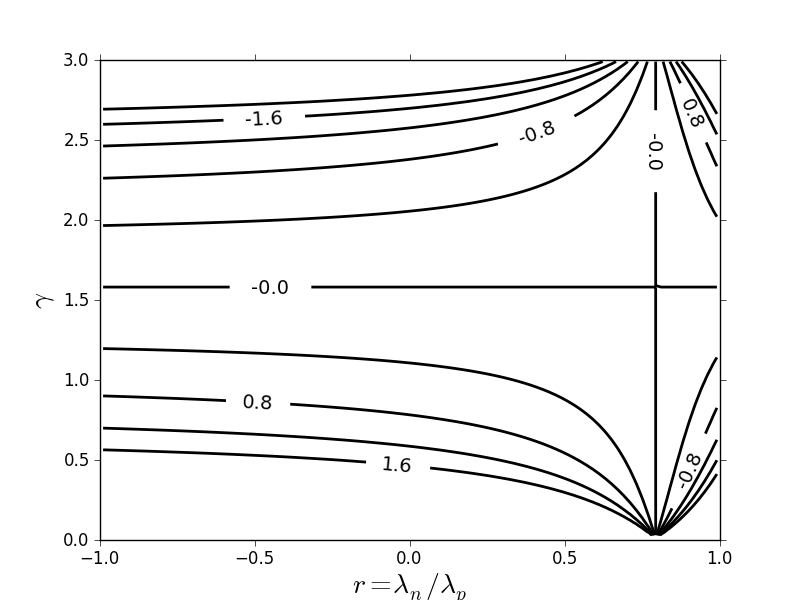}
\end{center}
\caption{
 Contour plot of $y_d^{\prime}/y_u^{\prime}$:
 We illustrate the contour for
 $y_d^{\prime}/y_u^{\prime} = \pm [0.0,\,0.4,\,0.8,\,1.2,\,1.6]$ which is given
 by \eqref{eq:yprime}.
}
\end{figure}

To discuss the favored region for the direct detection, we often use the
neutralino-proton cross section
\begin{equation}
 \sigma_p = \frac{4}{\pi}\left(\frac{m_{\chi}m_p}{m_{\chi} + m_p}\right)^2
  \lambda_p^2
\end{equation}
and the preferred regions of DAMA and CoGeNT signals are estimated to be
\begin{equation}
 1.5 \times 10^{-40}\,\mathrm{cm}^2 \lesssim \sigma_p^{\text{(DAMA)}} \lesssim
  3.0 \times 10^{-40}\,\mathrm{cm}^2,
\end{equation}
\begin{equation}
 5.0 \times 10^{-41}\,\mathrm{cm}^2 \lesssim \sigma_p^{\text{(CoGeNT)}} \lesssim
  8.0 \times 10^{-41}\,\mathrm{cm}^2
\end{equation}
for $m_{\chi} \simeq 8\,\mathrm{GeV}$ if there is no isospin violation
($\lambda_p = \lambda_n$).
Here we adopt the larger sodium quenching factor $Q_{Na} = 0.5 \pm 0.1$ so that the
DAMA preferred region include the dark matter mass $m_{\chi} \simeq 8\,\mathrm{GeV}$
\cite{Hooper:2010uy}\cite{Schwetz:2011xm}.
On the other hand, these parameter regions are already excluded by the experiments
such as CDMS-II and XENON100.
In particular, XENON100 have reported a severe constraint
\begin{equation}
 \sigma_p \lesssim 4.0 \times 10^{-42}\,\mathrm{cm}^2
\end{equation}
for $m_{\chi} \simeq 8\,\mathrm{GeV}$.
To consider the case with an isospin violation, we introduce the effective
cross section of neutralino and nucleus defined by
\begin{equation}
 \sigma_{\text{eff}}^{\text{SI}} \equiv \sum_i\frac{4\mu_{\chi i}^2}{\pi}p_i
  (\lambda_pZ + \lambda_n(A_i - Z))^2 \\
\end{equation}
where $i$ is summed over isotopes $A_i$ with fractional number abundance $p_i$.
This definition reflects the practical constituents of the detector with assumption
that only one of the elements dominates the contribution for the cross section.
The effect of the isospin violation can be extracted by comparing with
the cross section without isospin violation ($r = 1$):
\begin{equation}
 \sigma_{\text{eff}}^{\text{SI}} = K_r\sigma_{\text{eff}(r=1)}^{\text{SI}},\qquad
  K_r \equiv
  \frac{\sum_i\mu_{\chi i}^2p_i[(1-r)Z + rA_i]^2}{\sum_i\mu_{\chi i}^2p_iA_i^2}.
\end{equation}
Therefore, the favored region of the neutralino-proton cross section in the case
of isospin violation can be obtained by replacing $\sigma_p$ with $K_r\sigma_p$.

\begin{figure}[t]
\begin{center}
\includegraphics[width=10cm]{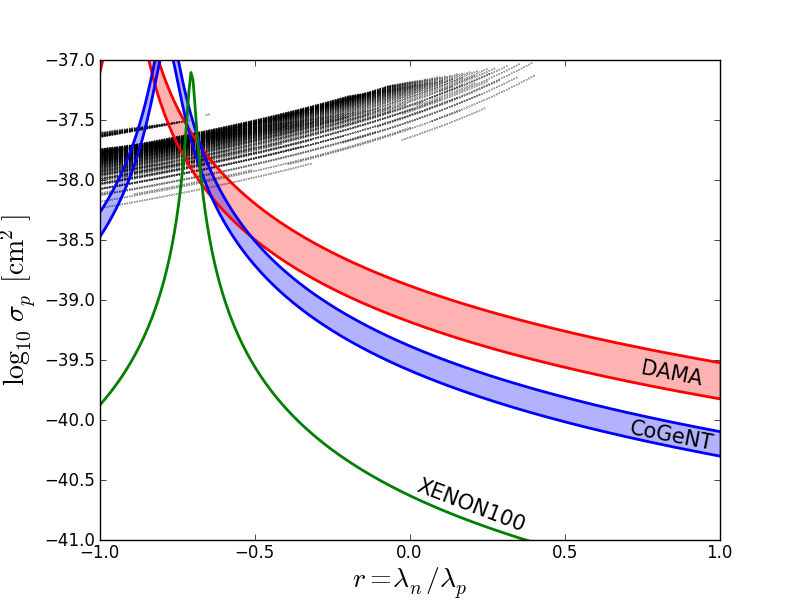}
\end{center}
\caption{
 Scatter plot of direct detection cross section and experimental preferred/excluded
 regions for $-1 \le r \le 1$.
 We set $m_{\chi} = 8\,\mathrm{GeV}$, $m_{\phi} = 150\,\mathrm{GeV}$,
 $\tan\beta_1 = 5$, $M_2 = 1\,\mathrm{TeV}$ and analyze the parameter space
 $\gamma = [0, 2\pi)$, $\mu_{12}, \mu_{21} = [100\,\mathrm{GeV}, 500\,\mathrm{GeV})$
 to draw the scatter plot.
 Yukawa couplings $y_u^{\prime}$ and $y_d^{\prime}$ are determined to generate the
 relic abundance of dark matter $\Omega_{\chi}h^2 \simeq 0.1$ with the help of
 \eqref{eq:sigmav} and \eqref{eq:yprime}
 In the scatter plot, we restrict $|y_u^{\prime}|,|y_d^{\prime}| \leq (4\pi)^{1/2}$.
}
\end{figure}

In Figure 2, we illustrate the scatter plot of direct detection cross section by varying
the ratio of neutron to proton coupling $r$.
Once we determine the ratio $y_d^{\prime}/y_u^{\prime}$ from \eqref{eq:yprime},
the Yukawa couplings $y_u^{\prime}$ and $y_d^{\prime}$ can be derived
by requiring to generate the relic abundance $\Omega_{\chi}h^2 \simeq 0.1$
as computed in previous subsection.
For evaluating $\sigma_p$, we adopt the mass of up-quark $m_u \sim 2\,\mathrm{MeV}$.
We also show the preferred region for DAMA and CoGeNT and constraint from XENON100
in figure 2.
As can be seen from the figure, this model can generate the cross section consistent
with the overlap region of DAMA and CoGeNT experiments.
Moreover, the constraint from XENON100 is weakened for $r \simeq -0.7$ and we can
find the parameter region consistent with these three experiments.
We have to mention that CDMS-II experiment still exclude the region favored by DAMA
and CoGeNT even if we introduce isospin violation.
This contradiction should be investigated carefully in the future experiments.
Recently, CRESST experiment has reported that they found signal consistent with
the light dark matter scenario \cite{Angloher:2011uu}.
We can also see that the 3$\sigma$ preferred region for CRESST result is compatible
with that of DAMA and CoGeNT if we take an isospin violation $r \simeq -0.7$.
\section{Collider signature}
\subsection{Constraints from precision experiments}
Although we have managed to explain the results from dark matter experiments
by supersymmetric four-Higgs doublet model, there can be some constraints from the
precision experiments for the parameters in this model.
In this subsection, we give some comments on the possible constraints from the current
experiments.

Since we now consider a very light neutralino with mass
$m_{\chi} \simeq 8\,\mathrm{GeV}$, we should care about the constraint from the
invisible decay rate of $Z$-boson.
The decay rate of $Z$-boson to a pair of the lightest neutralino can be written as
\begin{equation}
 \Gamma(Z^0 \to \chi_1^0\,\chi_1^0) = \frac{g^2}{96\pi c_W^2}M_Z
  \left(1 - \frac{4m_{\chi}^2}{M_Z^2}\right)^{3/2}
  \left(N_{15}^2 - N_{16}^2\right)^2
\end{equation}
which is similar to the MSSM case \cite{Komatsu:1986uz}\cite{Barbieri:1987hb}.
Therefore, arbitrary values of $N_{15}$ and $N_{16}$ given in \eqref{eq:n15} and
\eqref{eq:n16} with $|\mu_{12}|, |\mu_{21}| \gtrsim 100\,\mathrm{GeV}$ are
consistent with the constraint
$\Gamma(Z^0 \to \chi_1^0\,\chi_1^0) < 3\,\mathrm{MeV}$ as discussed in \cite{Fornengo:2010mk}.

When we introduce the extra Yukawa couplings, the severe constraints from flavor
physics generally appear as pointed out in the previous section.
Although the tree-level contribution for the flavor changing process is absent by
setting the Yukawa couplings of extra Higgs fields to the form given in
\eqref{eq:yukawa}, we should also care about the loop contributions mediated by
extra Higgs fields.
In this paper, we simply assume that the dangerous processes such as $b \to s\,\gamma$
are suppressed by the suitable choice of the off-diagonal elements of extra Yukawa
couplings and supersymmetry breaking parameters.
\subsection{$Wjj$ anomaly}
As discussed in the previous section, this model predicts light neutral extra Higgs
fields $\phi_{E1}^0$ and $\phi_{O1}^0$ to account for the relic abundance and
the preferred region for the direct detection of dark matter.
If such light fields exist with large Yukawa couplings, there is a possibility that
the current experiments already have an ability to discover (or exclude) these particles.
On the other hand, the CDF collaboration has reported that there is an excess
for the dijet mass distribution of the $W + j\,j$ event around 150 GeV.
This excess may be explained in our model if the final state two-jets are
mediated by the extra Higgs field.
Actually similar situation has been investigated in \cite{Cao:2011yt}
which introduce an additional quasi-inert Higgs doublet to the standard model.
Although this anomaly may not be true because no anomaly is seen by D0 experiment,
it is still important to investigate the possibility to test the model by the
existing experiments.

When we consider the dark matter relic abundance and the rate of direct detection,
the coupling constants are evaluated around the mass scale of the
neutralino $m_{\chi}$.
Now we have to incorporate the effect of the renormalization group flow
to calculate the production cross section of the extra Higgs.
Below the mass scale of the extra Higgs, only the $SU(3)_C$ gauge coupling
gives significant contribution for the running of the Yukawa couplings
$y_u^{\prime}$ and $y_d^{\prime}$.
So the renormalization group equations we should consider are given as
\begin{align}
 (4\pi)^2\frac{d}{d\ln\mu}\,y_{u/d}^{\prime} &= -8g_s^2y_{u/d}^{\prime}
  + \mathcal{O}\left(\frac{g_s^4y_{u/d}^{\prime}}{(4\pi)^2}\right), \\
 (4\pi)^2\frac{d}{d\ln\mu}\,g_s &= -\frac{23}{3}g_s^3
 + \mathcal{O}\left(\frac{g_s^5}{(4\pi)^2}\right)
\end{align}
at one-loop level.\footnote{
More precisely, we consider the running of the effective couplings
$\mathcal{A}_1^f(\bar{\chi}\chi)(\bar{f}f)
+ \mathcal{A}_2^f(\bar{\chi}\gamma^5\chi)(\bar{f}\gamma^5f)$
and take a matching at the scale $m_{\phi}$.
}
These equations can be solved to obtain
\begin{equation}
 y_{u/d}^{\prime}(m_{\phi}) =
  \left(\frac{g_s(m_{\phi})}{g_s(m_{\chi})}\right)^{24/23}
  \left[1 + \mathcal{O}\left(\frac{\Delta\alpha_s}{4\pi}\right)\right]
  \cdot y_{u/d}^{\prime}(m_{\chi})
\end{equation}
where $\Delta\alpha \equiv \alpha_s(m_{\chi}) - \alpha_s(m_{\phi})$ with
$\alpha_s \equiv g_s^2/(4\pi)$.
For $m_{\chi} = 8\,\text{GeV}$ and $m_{\phi} = 150\,\text{GeV}$,
we obtain $y_{u/d}^{\prime}(m_{\phi}) \simeq 0.7 \cdot y_{u/d}^{\prime}(m_{\chi})$
from $\alpha_s(m_{\chi}) \simeq 0.2$ and $\alpha_s(m_{\phi}) \simeq 0.1$.

If we accept the result from the CDF collaboration as the true signal of new particle,
their result suggests that the size of new physics contribution for the cross section of
$W + j\,j$ process is comparable with that from $W^+\,W^-$ production process.
Since $\sigma(p\,\bar{p} \to W^+\,W^-) \simeq 8.6\,\mathrm{pb}$ for
$\sqrt{s} = 1.96\,\text{TeV}$, it roughly require
the cross section for the process $p\,\bar{p} \to W^+\,\phi_{E1/O1}$ around
a few pb.

\begin{figure}[t]
\begin{center}
\includegraphics[width=10cm]{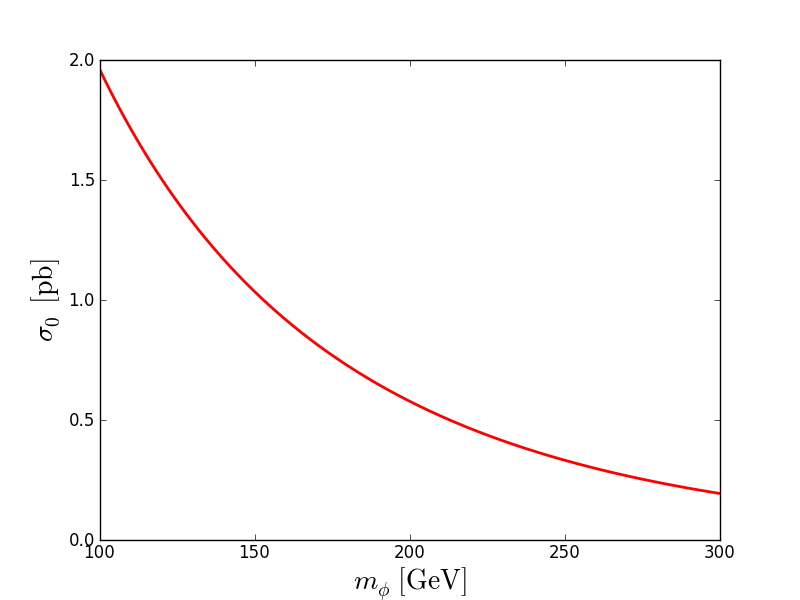}
\end{center}
\caption{
 Cross section $\sigma_0$ defined by \eqref{eq:sigma0} as a function of the mass
 of extra neutral Higgs $m_{\phi}$.
}
\end{figure}

To estimate the cross section, we assume that extra charged Higgs are much heavier than
$m_{\phi}$ and consider only the $t$-channel quark exchange contribution of
$\phi_{E1}$ and $\phi_{O1}$ final states for simplicity.
Note that the contribution from the resonant production of charged extra Higgs
$\phi^{\pm}$ is not so large because the branching ratio
$\mathrm{Br}(\phi^{\pm} \to W^{\pm}\,\phi^0)$ is very small compared with
$\mathrm{Br}(\phi^{\pm} \to j\,j)$ through large Yukawa coupling.
We write the cross section of this process as
\begin{equation}
 \sigma(p\,\bar{p} \to W^+\,\phi_{E1/O1}^0) \equiv
  \left(y_u^{\prime 2}c_{\gamma}^2 + y_d^{\prime 2}s_{\gamma}^2\right)\sigma_0
  \label{eq:sigma0}
\end{equation}
and $\sigma_0$ for various mass of extra Higgs $m_{\phi}$ is plotted in Figure 3.
Here we used LHAPDF \cite{Whalley:2005nh} to compute the cross section.
From this figure, we can see that required size of cross section can be obtained
by $\mathcal{O}(1)$ Yukawa couplings.
Of course, we have to carry out a Monte Carlo simulation to check whether this
model can generate a signal consistent with the experiment, but it is beyond the
scope of this paper.
Since this model also gives large cross section for the processes
$p\,\bar{p} \to Z^0\,\phi_{E1/O1}$ and $p\,\bar{p} \to \gamma\,\phi_{E1/O1}$,
precise measurements for these processes are important to test the model.
\section{Conclusion}
In this paper, we have considered supersymmetric four-Higgs doublet model to explain
the signals of dark matter direct detection from DAMA and CoGeNT experiments.
We have evaluated the relic abundance and dark matter-nucleus cross section for
the neutralino LSP scenario including the contributions from the extra Higgs
superfields.

We have shown that the preferred regions given by DAMA and CoGeNT can be
simultaneously explained by the supersymmetric four-Higgs doublet model
with the parameters which are consistent with the observed value of dark matter
relic abundance.
To explain the results of DAMA and CoGeNT, an isospin violation for the
neutralino-nucleus scattering is essential and we can achieve this situation
by adjusting the ratio of extra Yukawa couplings $y_d^{\prime}/y_u^{\prime}$.

The extra Higgs fields we introduced as an explanation of the light dark matter
are candidate for the solution of $Wjj$ anomaly reported by CDF.
We have calculated the cross section for the process
$p\,\bar{p} \to W^+\,\phi_{E1/O1}^0$ and the size of this cross section is
compatible with the experimental data.
\section*{Acknowledgments}
The author would like to thank N.~Maekawa, K.~Sakurai and N.~Nagata for useful
discussions and comments.
The author is supported by Grants-in-Aid for JSPS fellows.
This research was partially supported by the Grant-in-Aid for Nagoya Global
COE Program, ``Quest for Fundamental Principles in the Universe: from Particles
to the Solar System and the Cosmos'', from the MEXT of Japan.

\begin{thebibliography}{99}
\bibitem{Bernabei:2008yi}
  R.~Bernabei {\it et al.}  [DAMA Collaboration],
  Eur.\ Phys.\ J.\  C {\bf 56}, 333 (2008)
  [arXiv:0804.2741 [astro-ph]].
\bibitem{Aalseth:2010vx}
  C.~E.~Aalseth {\it et al.}  [CoGeNT collaboration],
  Phys.\ Rev.\ Lett.\  {\bf 106}, 131301 (2011)
  [arXiv:1002.4703 [astro-ph.CO]].
\bibitem{Angloher:2011uu}
  G.~Angloher {\it et al.},
  arXiv:1109.0702 [astro-ph.CO].
\bibitem{Aprile:2011hi}
  E.~Aprile {\it et al.}  [XENON100 Collaboration],
  arXiv:1104.2549 [astro-ph.CO].
\bibitem{Ahmed:2010wy}
  Z.~Ahmed {\it et al.}  [CDMS-II Collaboration],
  Phys.\ Rev.\ Lett.\  {\bf 106}, 131302 (2011)
  [arXiv:1011.2482 [astro-ph.CO]].
\bibitem{Bernabei:2010mq}
  R.~Bernabei {\it et al.},
  Eur.\ Phys.\ J.\  C {\bf 67}, 39 (2010)
  [arXiv:1002.1028 [astro-ph.GA]].
\bibitem{Ralston:2010bd}
  J.~P.~Ralston,
  arXiv:1006.5255 [hep-ph].
\bibitem{Nygren:2011xu}
  D.~Nygren,
  arXiv:1102.0815 [astro-ph.IM].
\bibitem{Blum:2011jf}
  K.~Blum,
  arXiv:1110.0857 [astro-ph.HE].
\bibitem{Foot:2008nw}
  R.~Foot,
  Phys.\ Rev.\  D {\bf 78}, 043529 (2008)
  [arXiv:0804.4518 [hep-ph]].
\bibitem{Khlopov:2008ki}
  M.~Y.~Khlopov,
  arXiv:0806.3581 [astro-ph].
\bibitem{Masso:2009mu}
  E.~Masso, S.~Mohanty and S.~Rao,
  Phys.\ Rev.\  D {\bf 80}, 036009 (2009)
  [arXiv:0906.1979 [hep-ph]].
\bibitem{Chang:2008gd}
  S.~Chang, G.~D.~Kribs, D.~Tucker-Smith and N.~Weiner,
  Phys.\ Rev.\  D {\bf 79}, 043513 (2009)
  [arXiv:0807.2250 [hep-ph]].
\bibitem{Cui:2009xq}
  Y.~Cui, D.~E.~Morrissey, D.~Poland and L.~Randall,
  JHEP {\bf 0905}, 076 (2009)
  [arXiv:0901.0557 [hep-ph]].
\bibitem{Kim:2009ke}
  Y.~G.~Kim and S.~Shin,
  JHEP {\bf 0905}, 036 (2009)
  [arXiv:0901.2609 [hep-ph]].
\bibitem{Bae:2010hr}
  K.~J.~Bae, H.~D.~Kim and S.~Shin,
  Phys.\ Rev.\  D {\bf 82}, 115014 (2010)
  [arXiv:1005.5131 [hep-ph]].
\bibitem{Feng:2011vu}
  J.~L.~Feng, J.~Kumar, D.~Marfatia and D.~Sanford,
  Phys.\ Lett.\  B {\bf 703}, 124 (2011)
  [arXiv:1102.4331 [hep-ph]].
\bibitem{Gao:2011ka}
  X.~Gao, Z.~Kang and T.~Li,
  arXiv:1107.3529 [hep-ph].
\bibitem{Bottino:2002ry}
  A.~Bottino, N.~Fornengo and S.~Scopel,
  Phys.\ Rev.\  D {\bf 67}, 063519 (2003)
  [arXiv:hep-ph/0212379].
\bibitem{Bottino:2003iu}
  A.~Bottino, F.~Donato, N.~Fornengo and S.~Scopel,
  Phys.\ Rev.\  D {\bf 68}, 043506 (2003)
  [arXiv:hep-ph/0304080].
\bibitem{Fornengo:2010mk}
  N.~Fornengo, S.~Scopel and A.~Bottino,
  Phys.\ Rev.\  D {\bf 83}, 015001 (2011)
  [arXiv:1011.4743 [hep-ph]].
\bibitem{Feldman:2010ke}
  D.~Feldman, Z.~Liu and P.~Nath,
  Phys.\ Rev.\  D {\bf 81}, 117701 (2010)
  [arXiv:1003.0437 [hep-ph]].
\bibitem{Kuflik:2010ah}
  E.~Kuflik, A.~Pierce and K.~M.~Zurek,
  Phys.\ Rev.\  D {\bf 81}, 111701 (2010)
  [arXiv:1003.0682 [hep-ph]].
\bibitem{Vasquez:2010ru}
  D.~A.~Vasquez, G.~Belanger, C.~Boehm, A.~Pukhov and J.~Silk,
  Phys.\ Rev.\  D {\bf 82}, 115027 (2010)
  [arXiv:1009.4380 [hep-ph]].
\bibitem{Vasquez:2011yq}
  D.~A.~Vasquez, G.~Belanger and C.~Boehm,
  arXiv:1108.1338 [hep-ph].
\bibitem{Escudero:2005hk}
  N.~Escudero, C.~Munoz and A.~M.~Teixeira,
  Phys.\ Rev.\  D {\bf 73}, 055015 (2006)
  [arXiv:hep-ph/0512046].
\bibitem{Gupta:2009wn}
  R.~S.~Gupta and J.~D.~Wells,
  Phys.\ Rev.\  D {\bf 81}, 055012 (2010)
  [arXiv:0912.0267 [hep-ph]].
\bibitem{Kanemura:2010pa}
  S.~Kanemura, T.~Shindou and K.~Yagyu,
  Phys.\ Lett.\  B {\bf 699}, 258 (2011)
  [arXiv:1009.1836 [hep-ph]].
\bibitem{Marshall:2010qi}
  G.~Marshall and M.~Sher,
  Phys.\ Rev.\  D {\bf 83}, 015005 (2011)
  [arXiv:1011.3016 [hep-ph]].
\bibitem{Haba:2011ra}
  N.~Haba and O.~Seto,
  Prog.\ Theor.\ Phys.\  {\bf 125}, 1155 (2011)
  [arXiv:1102.2889 [hep-ph]].
\bibitem{Aoki:2011yy}
  M.~Aoki, S.~Kanemura, T.~Shindou and K.~Yagyu,
  arXiv:1108.1356 [hep-ph].
\bibitem{Kanemura:2011fy}
  S.~Kanemura, E.~Senaha and T.~Shindou,
  Phys.\ Lett.\  B {\bf 706}, 40 (2011)
  [arXiv:1109.5226 [hep-ph]].
\bibitem{Hooper:2010uy}
  D.~Hooper, J.~I.~Collar, J.~Hall, D.~McKinsey and C.~Kelso,
  Phys.\ Rev.\  D {\bf 82}, 123509 (2010)
  [arXiv:1007.1005 [hep-ph]].
\bibitem{Schwetz:2011xm}
  T.~Schwetz and J.~Zupan,
  JCAP {\bf 1108}, 008 (2011)
  [arXiv:1106.6241 [hep-ph]].
\bibitem{Aaltonen:2011mk}
  T.~Aaltonen {\it et al.}  [CDF Collaboration],
  Phys.\ Rev.\ Lett.\  {\bf 106}, 171801 (2011)
  [arXiv:1104.0699 [hep-ex]].
\bibitem{Segre:2011tn}
  G.~Segre and B.~Kayser,
  arXiv:1105.1808 [hep-ph].
\bibitem{Cao:2011yt}
  Q.~H.~Cao, M.~Carena, S.~Gori, A.~Menon, P.~Schwaller, C.~E.~M.~Wagner and L.~T.~M.~Wang,
  JHEP {\bf 1108}, 002 (2011)
  [arXiv:1104.4776 [hep-ph]].
\bibitem{Masip:1995sm}
  M.~Masip and A.~Rasin,
  Phys.\ Rev.\  D {\bf 52}, 3768 (1995)
  [arXiv:hep-ph/9506471].
\bibitem{Bottino:2008xc}
  A.~Bottino, N.~Fornengo, G.~Polesello and S.~Scopel,
  Phys.\ Rev.\  D {\bf 77}, 115026 (2008)
  [arXiv:0801.3334 [hep-ph]].
\bibitem{Kolb:1990vq}
  E.~W.~Kolb and M.~S.~Turner,
  Front.\ Phys.\  {\bf 69} (1990) 1.
\bibitem{Pavan:2001wz}
  M.~M.~Pavan, I.~I.~Strakovsky, R.~L.~Workman and R.~A.~Arndt,
  PiN Newslett.\  {\bf 16}, 110 (2002)
  [arXiv:hep-ph/0111066].
\bibitem{Belanger:2008sj}
  G.~Belanger, F.~Boudjema, A.~Pukhov and A.~Semenov,
  Comput.\ Phys.\ Commun.\  {\bf 180}, 747 (2009)
  [arXiv:0803.2360 [hep-ph]].
\bibitem{Ohki:2008ff}
  H.~Ohki {\it et al.},
  Phys.\ Rev.\  D {\bf 78}, 054502 (2008)
  [arXiv:0806.4744 [hep-lat]].
\bibitem{Komatsu:1986uz}
  H.~Komatsu,
  Phys.\ Lett.\  B {\bf 177}, 201 (1986).
\bibitem{Barbieri:1987hb}
  R.~Barbieri, G.~Gamberini, G.~F.~Giudice and G.~Ridolfi,
  Phys.\ Lett.\  B {\bf 195}, 500 (1987).
\bibitem{Whalley:2005nh}
  M.~R.~Whalley, D.~Bourilkov and R.~C.~Group,
  arXiv:hep-ph/0508110.
\end{thebibliography}
\end{document}